\documentclass{emulateapj} 
 
\usepackage{natbib} 
\usepackage{hyperref} 
\usepackage{color} 
\usepackage{xspace} 
\shorttitle{Dry mergers and the evolution of massive early-type galaxies} 
\shortauthors{Sonnenfeld, Nipoti \& Treu} 
 
\def\ucsb{1} 
\def\bologna{2}

\def\mhalo{M_{\mathrm{h}}} 
\def\reff{R_{\mathrm{eff}}} 
\def\rein{R_{\mathrm{Ein}}} 
\def\ximin{\xi_{\mathrm{min}}} 
 
\def\Sref#1{Section~\ref{#1}\xspace} 
\def\Fref#1{Figure~\ref{#1}\xspace} 
\def\Tref#1{Table~\ref{#1}\xspace} 
\def\Eref#1{Equation~\ref{#1}\xspace}

\begin{document} 
 
\title{Purely dry mergers do not explain the observed evolution of massive early-type galaxies since $z\sim1$} 
\author{Alessandro~Sonnenfeld\altaffilmark{\ucsb}$^{*}$} 
\author{Carlo~Nipoti\altaffilmark{\bologna}} 
\author{Tommaso~Treu\altaffilmark{\ucsb}$^{\dag}$} 
%\author{Rapha\"el~Gavazzi\altaffilmark{\iap}} 
%\author{Sherry~H.~Suyu\altaffilmark{\ucsb,\kipac,\asiaa}} 
%\author{Philip~J.~Marshall\altaffilmark{\oxford}} 
%\author{Matthew~W.~Auger\altaffilmark{\datapeople}} 
%\author{the SL2S collaboration} 
 
% Auger, Sonnenfeld, Treu, Suyu: 
\altaffiltext{\ucsb}{Physics Department, University of California, Santa Barbara, CA 93106, USA}  
% Gavazzi, Brault: 
\altaffiltext{\bologna}{ 
Department of Physics and Astronomy, Bologna University, viale Berti-Pichat 6/2, 40127 Bologna, Italy} 
 
\altaffiltext{*}{{\tt sonnen@physics.ucsb.edu}} 
\altaffiltext{$\dag$}{{Packard Research Fellow}}

%------------------------------------------------------------------------------- 
 
\begin{abstract} 
Several studies have suggested that the observed size evolution of
massive early-type galaxies (ETGs) can be explained as a combination
of dry mergers and progenitor bias, at least since $z\sim1$. In this
paper we carry out a new test of the dry-merger scenario based on
recent lensing measurements of the evolution of the mass density
profile of ETGs. We construct a theoretical model for the joint
evolution of the size and mass density profile slope $\gamma'$ driven
by dry mergers occurring at rates given by cosmological
simulations. Such dry-merger model predicts a strong
  decrease of $\gamma'$ with cosmic time, inconsistent with the almost
  constant $\gamma'$ inferred from observations in the redshift range
  $0<z<1$. We then show with a simple toy model that a modest amount
of cold gas in the mergers -- consistent with the upper limits on
recent star formation in ETGs -- is sufficient to reconcile the model
with measurements of $\gamma'$.  By fitting for the amount of gas
  accreted during mergers, we find that models with dissipation are
  consistent with observations of the evolution in both size and
  density slope, if $\sim4$\% of the total final stellar mass arises
  from the gas accreted since $z\sim1$. Purely dry merger models are
  ruled out at $> 99\%$ CL.  We thus suggest a scenario where the
outer regions of massive ETGs grow by accretion of stars and dark
matter, while small amounts of dissipation and nuclear star formation
conspire to keep the mass density profile constant and approximately
isothermal.
\end{abstract} 
 
 
%------------------------------------------------------------------------------- 
 
\section{Introduction}\label{sect:intro} 
 One of the challenges faced by cosmological models is reproducing 
the observed population of massive quiescent galaxies across cosmic 
times.  In recent years, there has been growing interest in the 
problem posed by the apparent rapid size evolution of quiescent 
galaxies between $z\sim2$ and $z\sim1$ 
\citep[e.g.][]{Dad++05,Tru++06,vDo++08,Cas++11}. How much the evolution of the mass-size relation corresponds to a physical size 
growth of individual objects as opposed to progenitor bias is still 
under debate \citep{New++12b,Car++13,BNE13}. 
 
Theoretical studies aimed at matching the observed size evolution of
quiescent galaxies have focused on dissipationless (dry) mergers
\citep{NJO09,Nip++09,vdW++09,Hop++10d,Ose++12,Hil++13}, as the low
star formation rates measured in these galaxies leaves little room for
a significant occurrence of dissipative (wet) mergers.  The predicted
and observed merger rates in a dry-merger scenario, while still
insufficient to reproduce the size growth observed at $z\gtrsim1.5$,
seem to be able to account for the late ($z\lesssim1.5$) size
evolution of quiescent galaxies \citep{Nip++12,New++12b,Pos++13}.
In particular, \citet{Nip++12} have shown that, on average, the
  predictions of a purely dry merger model are marginally consistent
  with the observationally inferred evolution of the $M_*-\mhalo$ and
  $M_*-\reff$ relations in the redshift range $0\lesssim z \lesssim
  1.3$. Dry mergers, however, appear difficult to reconcile with the
tightness of the observed scaling relations
\citep{Nip++09,Nip++12,Sha++13}. It is not clear then if models based
purely on dry mergers can capture the relevant aspects of the
evolution of early-type galaxies (ETGs), or if additional physical
ingredients are required. In order to make progress, new observational
tests are needed.
 
We introduce in this paper a new test by adding to the size-evolution 
constraints the recent measurement of the evolution of the slope 
$\gamma'$ of the total density profile of massive ($M_* > 
10^{11}M_\odot$) ETGs in the range $0 < z < 1$. \citet{paperIV} show 
that ETGs increase in mass and size while keeping their density slope 
approximately constant and close to isothermal ($\gamma' \approx 
2$). By combining these two observational constraints we show that an 
evolution driven by purely dry mergers is ruled out, and some amount 
of dissipation is needed. 
 
The paper is organized as follows.  In \Sref{sect:dry} we construct a 
dry-merger evolutionary model and we compare with observations the 
evolution of $\gamma'$ for a sample of mock galaxies. In 
\Sref{sect:wet} we extend the model by including dissipation in the 
mergers and compare the predictions of this new model with 
observations. In \Sref{sect:popstudy} we quantify the amount of 
dissipation needed to fit both observational constraints and 
compare. We then discuss our results in 
\Sref{sect:discuss} and conclude in \Sref{sect:concl}. 
%------------------------------------------------------------------------------- 
 
\section{Dry mergers} 
\label{sect:dry} 
 
\subsection{Evolution in mass, size, and density slope}\label{sect:model} 
 
In the dry-merger scenario, galaxies increase their stellar and dark 
mass by accreting material from other galaxies. No new stars are 
generated during or after the merging process.  \citet{Nip++12} 
presented an analytic model, based on both cosmological and 
galaxy-merger $N$-body simulations, which allows to compute the 
dry-merging driven evolution of halo mass $\mhalo$, stellar mass 
$M_*$, effective radius $\reff$ and velocity dispersion, expected for 
spheroidal galaxies in $\Lambda$CDM cosmology. We refer the 
  reader to \citet{Nip++12} for a detailed description of the model: 
  here we just recall that the merger rate, as a function of $z$, 
  $\mhalo$ and merger mass ratio $\xi$, is the one measured in the 
  Millennium simulations \citep{Fak++10} and that the variations 
  in $\reff$ and $M_*$ are related by 
\begin{equation}\label{eq:drdm} 
\frac{\mathrm{d}\ln \reff}{\mathrm{d}\ln M_*}(\xi) = \left[2 - \frac{\ln{(1 + \xi^{2-\beta_R})}}{\ln{(1+\xi)}}\right], 
\end{equation} 
where $\beta_R$ is the logarithmic slope of the stellar mass-size 
  relation ($\reff\propto M_*^{\beta_R}$).  The model depends on few 
parameters (essentially $\beta_R$ and the minimum merger mass ratio 
$\ximin$) and on the stellar-to-halo mass relation (SHMR) used to 
associate halo and stellar masses. Here we adopt the model of 
\citet{Nip++12} with $\ximin=0.03$, $\beta_R=0.6$ and \citet{Lea++12} 
SHMR, but we verified that our results do not depend significantly on 
these choices. 
 
In this paper we extend the model by computing the change in the slope
$\gamma'$ of the total density profile to be compared with
measurements of the same quantity from the lensing and stellar
kinematics study of \citet{paperIV}. In practice, we need a formula
analogous to equation~(\ref{eq:drdm}), which gives
$\mathrm{d}\gamma'/\mathrm{d}\ln M_*$ expected for dry mergers as a
function of the merger mass ratio of $\xi$. For this purpose, we
  use a set of dissipationless binary-merger $N$-body simulations,
  which are described in \Sref{sect:simu}.  The analysis of these
  $N$-body calculations leads us to parameterize the change in
  $\gamma'$ resulting from mergers of mass ratio $\xi$ as
\begin{equation}\label{eq:dgdm} 
\frac{\mathrm{d}\gamma'}{\mathrm{d}\ln{M_*}}(\xi) = a\xi + b, 
\end{equation} 
with $a = 0.6$ and $b = -0.73$ (dashed line in \Fref{fig:xideltag}).
In practice, dry mergers make the density profile shallower and, for
the same amount of total accreted mass, minor mergers are more
effective at changing the density slope than major mergers.

  We note that throughout the paper $\xi$ indicates the {\it dark matter 
    mass ratio} between the satellite and the main galaxy. The
  corresponding {\it stellar mass ratio} is in general different from
  $\xi$, because $M_*/\mhalo$ depends on $\mhalo$. In our model when a
  halo of mass $\mhalo$ undergoes a merger with mass ratio $\xi$ the
  increase in dark matter mass is $\xi\mhalo$, and the increase in stellar mass
  is $\mathcal{R}_{\ast{\rm h}}\xi\mhalo$, where
  $\mathcal{R}_{\ast{\rm h}}$ is the ratio of stellar to dark matter mass of
  the satellite.  For given increase in stellar mass, the variation of
  $\reff$ and $\gamma'$ depends on $\xi$, but not on
  $\mathcal{R}_{\ast{\rm h}}$: this is justified because our $N$-body
  simulations indicate that the effect of varying
  $\mathcal{R}_{\ast{\rm h}}$ is small (see \Sref{sect:simu}).

\subsection{$N$-body simulations of binary dissipationless mergers}
\label{sect:simu} 

\begin{figure} 
\includegraphics[width=\columnwidth]{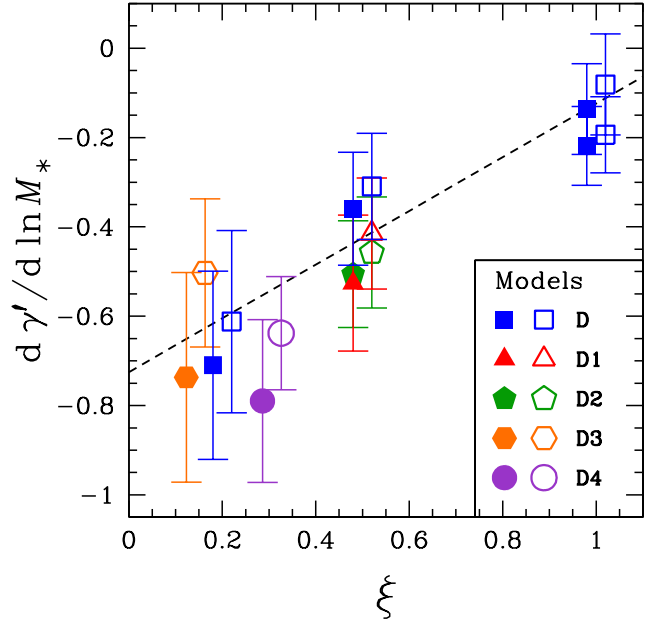} 
\caption{\label{fig:xideltag} Change in the density slope $\gamma'$
  per logarithmic unit of accreted stellar mass as a function of
  merger mass ratio $\xi$ in dissipationless binary-merger $N$-body
  simulations.  The different models are described in
  \Tref{tab:carlo}.  Filled symbols indicate head-on mergers while
  empty symbols refer to off-axis mergers (for the sake of clarity,
  the filled and empty points are shifted horizontally by -0.02 and
  0.02, respectively). The error bars account for projection
  effects. The dashed line is the linear best-fit to the set of models
  D.  In the case $\xi=1$ we consider two successive steps of a
    merger hierarchy: mergers of two D models (step 1; lower points
    the plot) and re-mergers of the remnant of step 1 with an
    identical system (upper points in the plot).}
\end{figure}

  In order to estimate $\mathrm{d}\gamma'/\mathrm{d}\ln M_*$ as a
  function of the merger mass ratio $\xi$ (see \Sref{sect:model}), we
  collect a set of dissipationless binary-merger $N$-body simulations
  by combining new simulations and simulations from previous works.
  In particular we take four simulations with $\xi=1$ and
  $\mhalo/M_*=49$ of \citet[][runs named 2D1ph, 2D1po, 4D1ph, 4D1po in
    table 2 of that paper]{NTB09} and two simulations with $\xi=0.2$
  and $\mhalo/M_*=49$ of \citet[][see section 3.3.2 in that
    paper]{Nip++12}. Our reference set of binary-merger simulations
  (named set D in \Tref{tab:carlo} and \Fref{fig:xideltag}) is
  supplemented by two new simulations with $\xi=0.5$ and
  $\mhalo/M_*=49$ (in all runs of this set the main galaxy and
  satellite have the same $\mhalo/M_*$).  In these simulations (except
  runs 4D1ph and 4D1po, which are re-mergers of runs 2D1ph and 2D1po;
  see \citealt{NTB09}) both the progenitor galaxies are represented by
  the two-component galaxy model D of \citet[][see table 1 of that
    paper]{NTB09}.  In all the runs of this set the ratio between the
  effective radius of the satellite and of the main galaxy is
  $\xi^{0.6}$ and the orbits are parabolic: some encounters are
  head-on ($r_{\rm peri}=0$), others are off-axis ($r_{\rm
    peri}/r_{\rm vir}\simeq 0.2$ for $\xi=0.2$ and $\xi=0.5$; see
  \cite{NTB09} for the orbital parameters of the $\xi=1$ runs). Here
$r_{\rm peri}$ is the pericentric radius and $r_{\rm vir}$ is the
virial radius of the main halo.
 
\begin{deluxetable*}{lccccccc} 
\tablewidth{0pt} 
\tablecaption{Paramaters of galaxy models in dissipationless binary-merger $N$-body simulations. } 
\tablehead{ 
\colhead{Set}   
& \colhead{$(\mhalo/M_*)_1$}   
& \colhead{$C_{1}$}   
& \colhead{$(r_s/\reff)_1$} 
& \colhead{$(\mhalo/M_*)_2$}   
& \colhead{$C_{2}$}   
& \colhead{$(r_s/\reff)_2$} 
} 
\startdata  
D   & 49 & 8.0  & 11.6 & 49 & 8.0  & 11.6 \\ 
D1   & 49 & 5.0  & 11.6 & 49 & 5.0  & 11.6 \\ 
D2   & 49 & 8.0  & 6.0 & 49 & 8.0  & 6.0 \\ 
D3   & 49 & 8.0  & 11.6 & 35 & 8.5  & 8.8 \\ 
D4   & 49 & 8.0  & 11.6 & 75 & 8.5  & 15.0  
\enddata 
\tablecomments{Set: name of the simulation set.  $C$: NFW 
  concentration. $r_s$: NFW scale radius. $\reff$: effective radius. $\mhalo$: total dark-matter 
  mass. $M_*$: total stellar mass. Subscript $1$ is for the main galaxy, subscript $2$ for the satellite.} 
\label{tab:carlo} 
\end{deluxetable*}

%which parameter has the larges impact on the change in $\gamma'$. 
%We consider equal mass ($\xi=1$), as well as 1:2 ($\xi=0.5$) and 1:5 ($\xi=0.2$) mergers, extending the simulations of \citet{NTB09} and \citet{Nip++12} with a series of model galaxies described in \Fref{fig:} 
 
%: in particular we consider the equal mass ($\xi=1$) merger 
%hierarchies of \citet{NTB09} and the minor ($\xi=0.2$) merger 
%simulations of \citet{Nip++12}.   
 
In order to minimize systematic errors, we measure $\gamma'$ in
simulated galaxies with the same method used in observations.  In
particular, the density slope $\gamma'$ measured by \citet{paperIV} is
obtained by fitting a power-law $\rho\propto r^{-\gamma'}$ to the
luminosity-weighted line-of-sight velocity dispersion within a
circular aperture of radius $\reff/2$ and to the total projected mass
within a cylinder of radius $\rein$.  For a given lens the value of
the Einstein radius $\rein$ depends on the distance of the lensed
background source: $\rein$ increases for increasing source redshift
$z_s$.  Typical strong lenses such as those of the SLACS
\citep{Aug++09a} or SL2S \citep{paperIII} surveys have Einstein radii
not too different from their effective radii. \citet{paperIV} showed
that measurements of $\gamma'$ with lensing and stellar dynamics are
very stable against variations of the ratio $\rein/\reff$.  When
measuring $\gamma'$ in simulated galaxies we will always assume
$\rein=\reff$, as varying the ratio $\rein/\reff$ has little impact on
the measured $\gamma'$.

  For our reference set of $N$-body simulations (set D)
  $\mathrm{d}\gamma'/\mathrm{d}\ln M_*$ as a function of the merger
  mass ratio $\xi$ is fitted by \Eref{eq:dgdm} with $a = 0.60 \pm
  0.19$ and $b = -0.73 \pm 0.13$. The evolution in $\gamma'$ is
  obtained by fixing the parameters $a$ and $b$ in \Eref{eq:dgdm} to
  their best-fit values. We verified that our results are robust
  against variations of $a$ and $b$ within the measured
  uncertainties. Of course, it is important to verify whether the
  adopted formula for $\mathrm{d}\gamma'/\mathrm{d}\ln M_*(\xi)$ is
  also robust against variation of the parameters characterizing the
  galaxy models. For this purpose we ran eight additional simulations
with the same orbital and galaxy parameters as the corresponding
simulations of set D, but changing the concentration $C\equiv r_{\rm
  vir}/r_{\rm s}$, where $r_{\rm s}$ is the Navarro Frenk and White
\citep[NFW][]{NFW97} scale radius, the stellar-to-halo mass ratio
$M_*/\mhalo$ and the ratio $r_{\rm s}/\reff$ of the progenitor
galaxies.  The values of the parameters of these additional sets of
simulations (named D1, D2, D3 and D4) are reported in \Tref{tab:carlo}
and are chosen to span the range of values expected for real
galaxies. We note that in all cases the progenitors have $\gamma'$ in
the range $1.97\lesssim\gamma'\lesssim 2.03$. The results of the runs
D1, D2, D3 and D4 are very similar to those of the corresponding runs
D (see \Fref{fig:xideltag}). Thus we conclude that our adopted formula
is robust with respect to variations in the properties of the host
galaxy and its satellite within realistic ranges.  
 
All the binary-merger $N$-body simulations were run with the 
  parallel $N$-body code FVFPS \citep[Fortran Version of a Fast 
    Poisson Solver;][]{LNC03,NLC03}.  The parameters of the 
  simulations with $\xi=1$ are given in \citet{NTB09}.  In the runs 
  with $\xi<1$ we adopted the following values of the code parameters: 
  minimum value of the opening parameter $\theta_{\rm min}=0.5$ and 
  softening parameter $\varepsilon=0.04\reff$, where $\reff$ is the 
  initial effective radius of the main galaxy. The time-step $\Delta 
  t$, which is the same for all particles, is allowed to vary 
  adaptively in time as a function of the maximum particle density 
  $\rho_{\rm max}$: in particular, we adopted $\Delta t=0.3/(4\pi G 
  \rho_{\rm max})^{1/2}$. The initial conditions of the new 
  simulations are realized as in \citet{NTB09}, but with dark matter 
  particles twice as massive as the stellar particles.  The total 
  number of particles used in each simulation is in the range 
  $1.6-3.4\times10^6$. 
 
In all the simulations used in this work the galaxy collision is
followed up to the virialization of the resulting stellar system. We
define the merger remnant as the systems composed by the bound stellar
and dark matter particles at the end of the simulation.  The intrinsic
and projected properties of the progenitors and of the merger remnants
are determined as in \citet{NTB09}, with the exception of $\gamma'$,
which, as pointed out above, is computed with the same procedure used
for observed lenses by \citet{paperIV}.

%------------------------------------------------------------------------------- 
 
\subsection{The model sample}\label{sect:sample}

Our goal is to follow the evolution of a sample of model galaxies 
between $z=1$ and $z=0$, matching the characteristic of the sample 
observed by \citet{paperIV}. We thus consider $N_{\mathrm{gal}}=1000$ 
objects with $\log{M_*}$ drawn from a Gaussian with mean $\mu_{*} = 
11.5$ and dispersion $\sigma_{*}=0.3$. The starting point of the 
evolutionary tracks of all galaxies is fixed at $z=0.3$, which is the 
redshift for which observations of $\gamma'$ are most robust. 
Effective radii are drawn from the mass-size relation measured by 
\citet{New++12b}. Halo masses are assigned with the same SHMR used in 
the galaxy evolution model described in \Sref{sect:model}.   
For fixed $M_*$, $\mhalo$ and $\reff$, the value of $\gamma'$ is not uniquely determined as this depends on additional parameters, such as the orbital anisotropy and the concentration of the dark matter halo. 
The 
initial values of $\gamma'$ are then drawn from the 
distribution measured by \citet{paperIV}. Once the initial values are set, $M_*$, $\mhalo$, 
$\reff$ and $\gamma'$ are evolved according to our model as described 
in \Sref{sect:model}. 
 
  Roughly half of the accreted stellar mass and the corresponding
  change in $\gamma'$ is due to mergers with $\xi < 0.2$. Since $\xi =
  0.2$ is the smallest mass ratio we consider in our $N$-body
  simulations, our predictions on the evolution of $\gamma'$ for the
  sample of mock galaxies relies in part on an extrapolation of
  \Eref{eq:dgdm}.  We verified that even in the extreme case in which
  the function $d\gamma'/d\ln{M_*}(\xi)$ flattens abruptly below $\xi
  = 0.2$ the conclusions of our analysis do not change.

%------------------------------------------------------------------------------- 
 
\subsection{Comparison with observations}\label{sect:comparison} 
 
The measurements by \citet{paperIV} constrain the parameter $\gamma'$ 
in ETGs as a function of their redshift, stellar mass and half-light 
radius.  The mean change of $\gamma'$ with one of these parameters and 
others fixed is measured to be 
\begin{equation}\label{eq:pgpz} 
\frac{\partial \gamma'}{\partial z} =-0.31\pm0.10, 
\end{equation} 
\begin{equation}\label{eq:pgpm} 
\frac{\partial \gamma'}{\partial \log{M_*}} = 0.40\pm0.16, 
\end{equation} 
\begin{equation}\label{eq:pgpr} 
\frac{\partial \gamma'}{\partial \log{\reff}} = -0.76\pm0.15. 
\end{equation} 
According to the formalism introduced in \citet{paperIV}, the observed change of $\gamma'$ with redshift for a galaxy with a mass growth rate $\mathrm{d}\log{M_*}/\mathrm{d}z$ and a size growth rate of $\mathrm{d}\log{\reff}/\mathrm{d}z$ is 
\begin{equation}\label{eq:lagrange} 
\frac{\mathrm{d}\gamma'}{\mathrm{d}z} =  
\frac{\partial \gamma'}{\partial z} + \frac{\partial \gamma'}{\partial \log M_*}\frac{\mathrm{d}\log M_*}{\mathrm{d}z} + \frac{\partial\gamma'}{\partial \log{\reff}}\frac{\mathrm{d}\log{\reff}}{\mathrm{d}z}. 
\end{equation} 
The quantities $\mathrm{d}\log{M_*}/\mathrm{d}z$ and 
$\mathrm{d}\log{\reff}/\mathrm{d}z$ are not constrained by the 
observations, but are directly provided by our model for the dry 
merger evolution of galaxies. 
%, therefore it is straightforward to 
%compute the variation in $\gamma'$ consistent with observations. 
 
An implicit assumption of \Eref{eq:lagrange} is that the observed 
trends of $\gamma'$ with stellar mass and size are determined uniquely 
by the intrinsic evolution of galaxies, and not by the appearance of 
new objects with time.  This is a reasonable approximation, given that 
the total number density of quiescent galaxies has little evolution 
since $z\sim1$ \citep{Cas++13}, particularly at the large masses of our 
sample \citep{Ilb++13}.

\begin{figure} 
\includegraphics[width=\columnwidth]{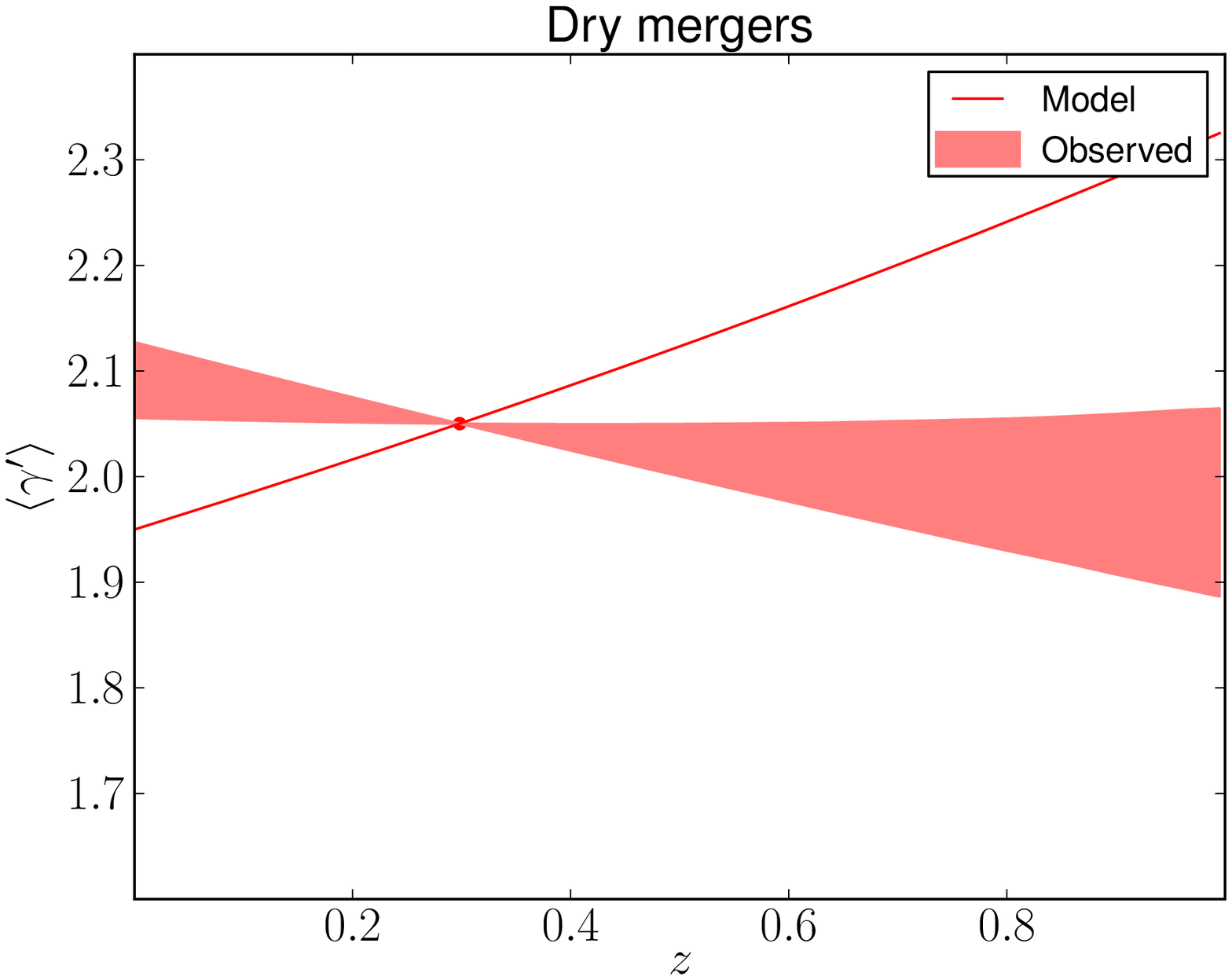} 
\includegraphics[width=\columnwidth]{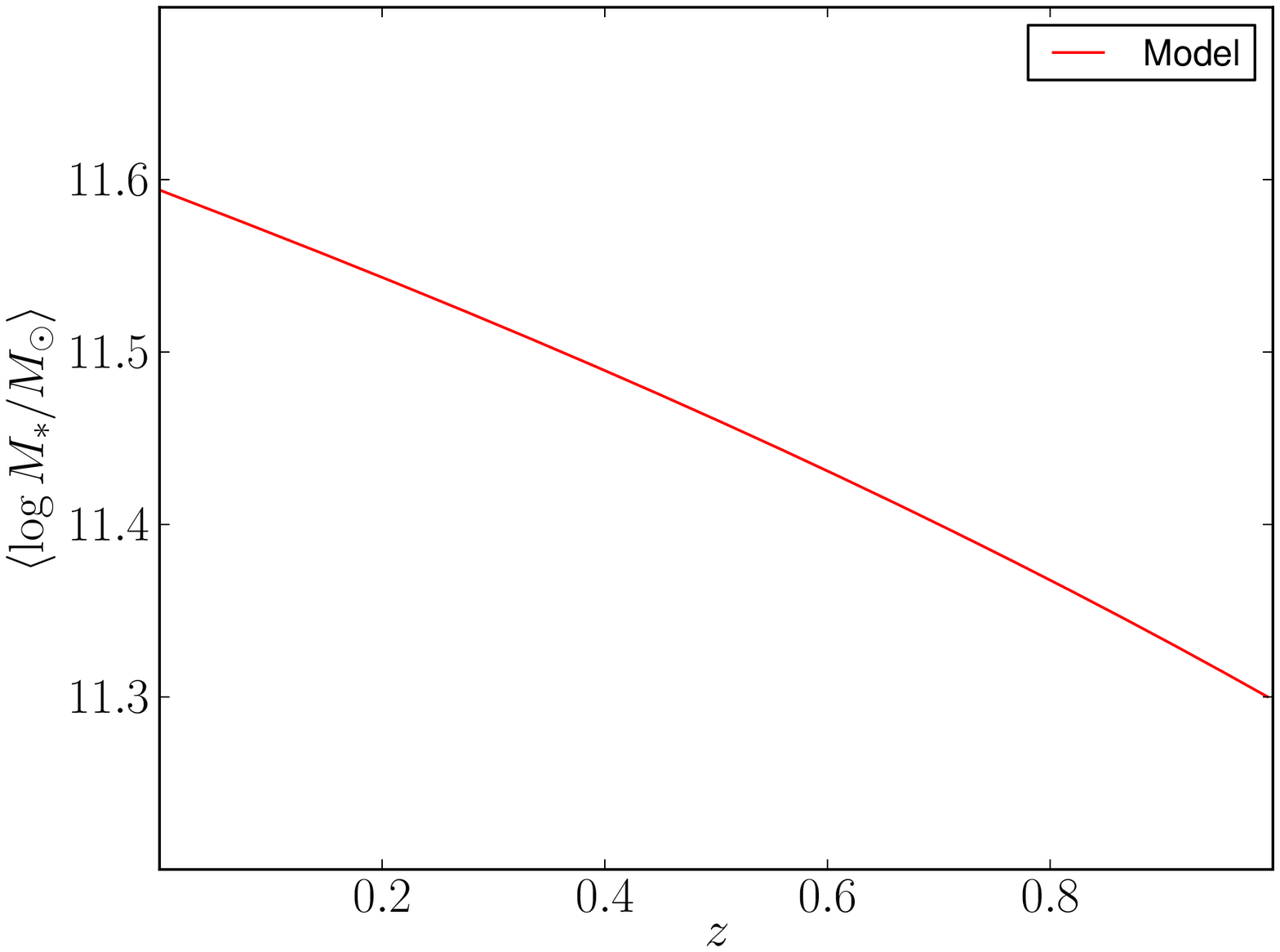} 
\caption{\label{fig:modelcB}  
{\em Top panel.}
{\em Solid line:} Density slope $\gamma'$, averaged over the mock galaxy population described in \Sref{sect:sample}, as a function of redshift. 
%The SHMR used is the one of \citet{Beh++10}. %The minimum mass ratio for mergers is $\xi_{\mathrm{min}} = 0.03$. 
{\em Shaded region:} 68\% confidence region for the observed change in $\gamma'$ for a population of galaxies with the same mass and size growth rate as the model one. 
{\em Bottom panel.} Average stellar mass of the mock galaxy sample as a function of redshift.
}
\end{figure}

\Fref{fig:modelcB} shows the model evolution of the density slope 
averaged over the sample of $1000$ galaxies, $\left<\gamma'\right>$. 
The mean change in $\gamma'$ with redshift for the sample average is 
$\mathrm{d}\gamma'/\mathrm{d}z = 0.33$, and the scatter over the 
sample is $\sigma_{d\gamma'/dz} = 0.15$. The average mass and size 
growth rates are $d\log{M_*}/dz = -0.27$ (galaxies roughly double in 
stellar mass from $z=1$ to $z=0$) and $d\log{\reff}/dz = -0.36$ 
respectively. The figure also shows the observed mean change in 
$\gamma'$ calculated following \Eref{eq:lagrange}. This is 
$d\gamma'/dz = -0.13\pm0.12$. The key result is that predicted and 
observed evolution in $\gamma'$ differ significantly. %and the model is 
%ruled out at more than 99\%CL. 

\Eref{eq:lagrange} provides an efficient way to quickly compare model predictions with observations.  
However, in this context the ``observed'' evolution of $\gamma'$ is really a combination of observed quantities (the partial derivatives) and model predictions ($d\log{M_*}/dz$ and $d\log{\reff}/dz$). 
A more direct evaluation of the goodness of the model is obtained by comparing models to the observables, i.e. the partial derivatives. 
Such a comparison is done in \Sref{sect:popstudy}. 
Here we simply point out a discrepancy between the predicted evolution of $\gamma'$ and observed data, the significance of which will be discussed later.

%------------------------------------------------------------------------------- 
 
\section{Wet mergers}\label{sect:wet}

The above analysis is based on the assumption that the growth of 
galaxies is a result of purely dry mergers.  In practice, mergers 
between galaxies are expected to involve the accretion of gas, which 
can radiate away energy and sink to the central parts of the main 
galaxy, eventually leading to star formation episodes.  This infall of 
gas can alter the density profile of the accreting galaxy, making it 
steeper.  Thus introducing dissipation in our model should help 
reproduce the observed evolution of $\gamma'$.  
 
Following the spirit of our approach we introduce dissipation using a 
simple toy model. In spite of its simplicity this approach allows us to 
isolate cleanly the effect of dissipation and estimate whether this 
solution can work at all. Thus it should provide a very good 
complement to hydrodynamic cosmological simulations which are just 
starting to achieve the resolution to model the internal structure of 
ETGs \citep{Fel++10,Ose++12,JNO++12,Rem++13,Dub++13}. 
 
We wish to test whether dissipation {\it can} work and therefore we 
consider a plausible yet somewhat extreme model which maximizes the 
effects on the mass profile.  In practice, we assume that a small 
  fraction of the baryonic mass of the merging satellite is cold gas, 
  which, in the merging process, falls exactly to the center of the 
  galaxy and forms stars. We calculate the response of the mass 
  distribution of the galaxy to the infall of cold gas following the 
  adiabatic contraction recipe of \citet{Blu++86}, which is stronger 
  than more recent ones based on numerical simulations 
  \citep[e.g.][]{Gne++04}. 
 
The galaxies are modeled as spherical de Vaucouleurs stellar bulges and 
a dark matter halo with an NFW profile. 
The ratio between halo mass and stellar 
mass is $M_h/M_* = 50$, and the ratio between the scale radius of the 
NFW profile and the effective radius of the stellar component is 
$r_s/R_{\mathrm{eff}}=10$. For given infalling mass we first calculate 
the new distribution of stellar and dark matter mass following 
adiabatic contraction. Then we calculate the stellar half-light 
radius. Finally we calculate $\gamma'$ consistently with lensing and 
dynamics measurements. 
 
\begin{figure} 
\includegraphics[width=\columnwidth]{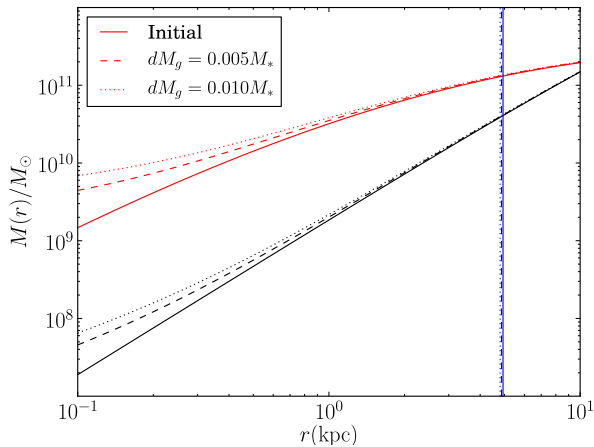} 
\caption{\label{fig:Mwet}  
Stellar (red) and dark matter (black) mass profiles for a model galaxy, before and after the infall of gas at the center and the subsequent adiabatic contraction. 
The blue vertical lines indicate $r=R_{\mathrm{eff}}$.
}
\end{figure}

\begin{figure}
\includegraphics[width=\columnwidth]{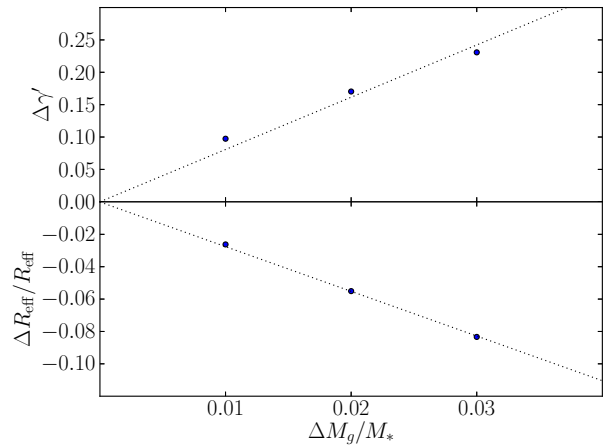}
\caption{\label{fig:linfit}
{\em Top panel:} Change in the density slope as a function of infalling gas (blue dots), and its best linear fit of \Eref{eq:wet_linfit} (dashed line).
{\em Bottom panel:} Logarithmic change in the effective radius as a function of infalling gas (blue dots), and its best linear fit of \Eref{eq:reff_linfit} (dashed line).
} 
\end{figure}

The change in $\reff$ and $\gamma'$ is caused by both the addition of 
new material at $r=0$ and by the subsequent contraction of the 
preexisting mass. The two effects have a comparable impact on $\gamma'$.
\Fref{fig:Mwet} shows the original mass distribution as well as 
the one following gas infall and adiabatic contraction for a typical 
system. The relation between the change in $\gamma'$ due to and the 
accreted gas mass $dM_g$ can be fitted with a linear relation 
\begin{equation}\label{eq:wet_linfit} 
\frac{\mathrm{d}\gamma'}{\mathrm{d}M_g} = \frac{c}{M_*}, 
\end{equation} 
with $c = 7.9$ (see \Fref{fig:linfit}). 
The exact value of $c$ depends on the properties of the main galaxy. However, for the same parameters explored in \Sref{sect:model} and summarized in \Tref{tab:carlo} the variation of $c$ is smaller than 10\%. 
% 
%\begin{figure} 
%\includegraphics[width=\columnwidth]{wet_gammap.eps} 
%\caption{\label{fig:wetmodel}  
%Change in the density slope $\gamma'$ due to the infall of gas to the galaxy center and subsequent adiabatic contraction, as a function of the ratio between the accreted gas mass and the stellar mass of the galaxy. 
%The different models correspond to the same properties of the main galaxy listed in \Tref{tab:carlo}. 
%j{\em Dashed line:} best linear fit. 
%} 
%\end{figure} 
 
This ingredient is then added to the model describing the evolution of 
$\gamma'$ in the dry merger case. The accreted gas mass is 
  assumed to be a fraction $f_g$ of the accreted stellar mass, so 
  $dM_g = f_gdM_*$: for simplicity we assume that $f_g$ is independent 
  of redshift, stellar and halo mass. Then \Eref{eq:dgdm} is modified 
to 
\begin{equation} 
\frac{\mathrm{d}\gamma'}{\mathrm{d}M_*}(\xi) = \frac{1}{M_*}\left[a\xi + b + cf_g\right]. 
\end{equation} 
The effect of the infall of gas on the effective radius is quantified as 
\begin{equation}\label{eq:reff_linfit} 
\frac{\mathrm{d}\log{\reff}}{\mathrm{d}\log{M_*}} = -2.8f_g,
\end{equation} 
which is a measure of the reduced increase of $\reff$ due to
  dissipation (see also \citealt{CLV07}).  This term will be added to
\Eref{eq:drdm} when calculating the size evolution in the model with
wet mergers.
 
\begin{figure} 
\includegraphics[width=\columnwidth]{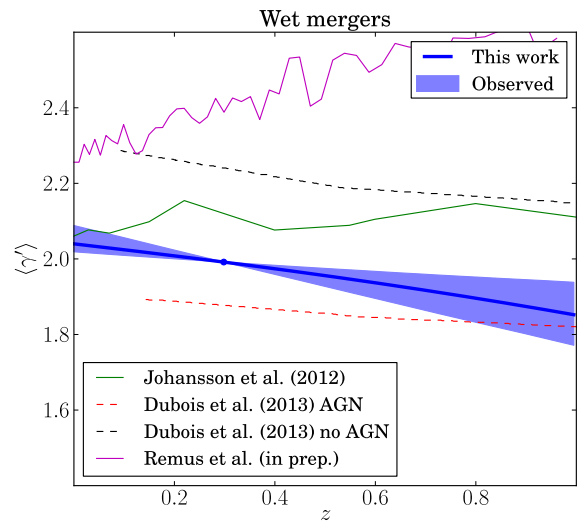} 
\caption{\label{fig:gammap_evol_wwet}  
{\em Solid line:} Density slope $\gamma'$, averaged over the mock galaxy population described in \Sref{sect:sample} and evolved taking into account the effects of wet mergers assuming $f_g = 0.1$. 
%The SHMR used is the one of \citet{Beh++10}.  
{\em Shaded region:} 68\% confidence region for the observed change in $\gamma'$ for a population of galaxies with the same mass and size growth rate as the model one. 
For comparison, we plot the average $\gamma'$ of galaxies from cosmological simulations of \citet{JNO++12}, \citet{Dub++13} and Remus et al. (in prep.). 
} 
\end{figure}

We set $f_g = 0.1$ and calculate the evolution of the population 
average of $\gamma'$ analogously to \Sref{sect:comparison}. Results 
are plotted in \Fref{fig:gammap_evol_wwet}, together with the observed 
evolution calculated with \Eref{eq:lagrange}. The average change in 
$\gamma'$ is now $\mathrm{d}\gamma'/\mathrm{d}z = -0.15$ with a 
scatter of $\sigma_{d\gamma'/dz} = 0.03$.  Note that this new model 
modifies also the interpretation of the observational results owing to 
the slightly smaller theoretical size growth entering 
\Eref{eq:lagrange}. The key result is that, by introducing a 
reasonable amount of dissipation, model predictions and observations 
are now in good agreement. 

\begin{figure} 
\includegraphics[width=\columnwidth]{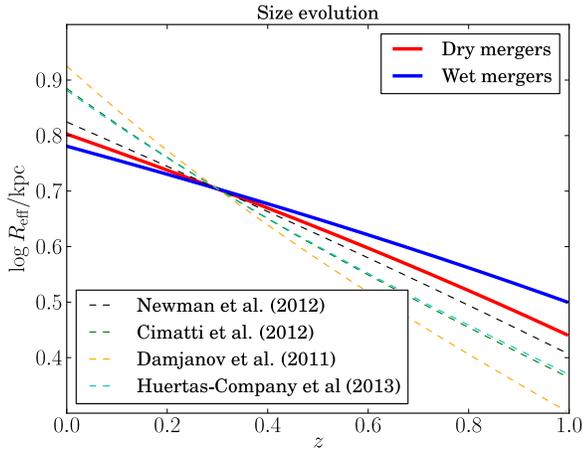} 
\caption{\label{fig:size}  
{\em Solid lines:} redshift evolution of the mock sample average of the effective radius in the dry (red) and wet (blue) merger case. A gas fraction $f_g = 0.1$ is assumed in the latter case. 
{\em Dashed lines:} observed size growth we infer from the best-fit size evolution measurements by \citet{Dam++11}, \citet{New++12b}, \citet{Cim++12}, \citet{Hue++13} assuming no progenitor bias. 
} 
\end{figure}

So far we have focused our attention to the evolution of the density slope $\gamma'$. 
There is another important piece of observations that a successful model of galaxy evolution needs to reproduce: the size evolution. 
We want to verify whether the two models considered so far predict a size growth consistent with observations. 
This is done in \Fref{fig:size}, where we plot $\reff$ as a function of $z$ 
for the sample average in both the dry and wet merger model, together 
with the observed average size evolution of galaxies with the same 
mass as the model average, for various literature measurements and 
assuming no progenitor bias.  Most measurements imply a stronger size 
evolution than our model predictions for both the dry and wet merger 
case, the discrepancy being worse for the wet merger model. 
Adding dissipational effects then helps matching $\gamma'$ observations, but increases the tension with size evolution data. 
Our knowledge of the size distribution of massive ETGs at $z < 1$ therefore should rule out models with too much dissipation. 
%Presumably, there is an optimal level of the parameter $f_g$ for which both model size and density slope can be reconciled with observations. 
In the following section we determine how much disspation is needed, if at all, to best match both sets of observables. This inference will also allow us to perform model selection, i.e to compare how well the purely dry merger scenario compares with the wet merger one. 
%In order to estabilish how much dissipation is required for an optimal  

%------------------------------------------------------------------------------ 
\section{Constraining the amount of dissipation}\label{sect:popstudy} 
 
\begin{figure} 
\includegraphics[width=\columnwidth]{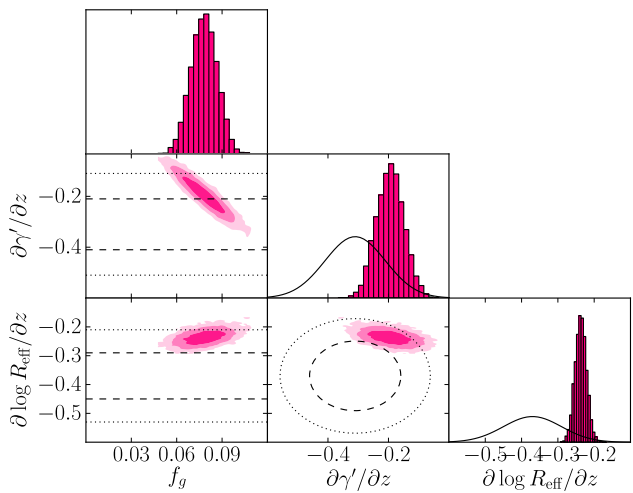} 
\caption{\label{fig:post}  
{\em Filled contours:}  
Posterior probabiliy distribution of the parameters describing the mock population of ETGs, projected on the space defined by the gas fraction, the dependence of $\gamma'$ on redshift, and the dependence of $\reff$ on redshift. 
{\em Dashed and dotted lines:} 68\% and 95\% enclosed probability of the observed redshift dependence of $\gamma'$, from \citet{paperIV}, and of the effective radius, obtained by combining measurements by \citet{New++12b}, \citet{Dam++11}, \citet{Cim++12} and \citet{Hue++13}, as explained in the text. 
} 
\end{figure}

As shown in \Sref{sect:wet}, the infall of cold gas and the 
  subsequent adiabatic contraction help reconcile the predicted 
  evolution of the density profile with observations.  At the same 
  time however adiabatic contraction leads to a decrease of the 
  effective radius, such that models with too much dissipation are in 
  tension with size evolution measurements.  The importance of these 
  two effects increases with increasing gas fraction, which we 
  parameterize with $f_g$.  We wish to estabilish which values of 
  $f_g$ provide the best match to the data, including both density 
  slope and size measurements.  We do this by generating mock 
  populations of ETGs, evolved with different values of $f_g$, and by 
  comparing scaling relations of $\gamma'$ and $\reff$ with 
  observations.  Our mock population is built by picking, for each one 
  of the $N_{\mathrm{gal}}=1000$ mock galaxies described in 
  \Sref{sect:sample}, a random snapshot on its evolutionary 
  track. This results in a set of galaxies uniformly distributed in 
  redshift in the interval $0 < z < 1$.  Sizes and density slopes of these mock galaxies will 
  depend on the amount of dissipation allowed by the model, 
  parameterized by $f_g$.  For fixed $f_g$, we can infer how the 
  average density slope of the mock population scales with redshift, 
  stellar mass and effective radius by measuring 
  $\partial\gamma'/\partial z$, $\partial\gamma'/\partial \log{M_*}$, 
  $\partial\gamma'/\partial \log{\reff}$ with the same method used by 
  \citet{paperIV}.  Similarly, we can measure how the average 
  effective radius scales with redshift and stellar mass. We assume 
  the following relation: 
\begin{eqnarray} 
\left<\log{\reff}\right> = & & \log{R_0} + \frac{\partial \log{\reff}}{\partial z} (z - 0.3) + \nonumber \\
& & + \frac{\partial\log{\reff}}{\partial\log{M_*}}(\log{M_*} - 11). 
\end{eqnarray} 
  We then fit for $f_g$ by comparing the model partial derivatives 
  of $\gamma'$ and $\reff$ with the values measured by \citet{paperIV} 
  (\Eref{eq:pgpz}, \Eref{eq:pgpm}, \Eref{eq:pgpr}) and in size 
  evolution studies.  The redshift dependence of the average effective 
  radius, $\partial \log{\reff}/\partial z$, has been measured by different 
  authors. As \Fref{fig:size} shows, there is some scatter between the 
  reported values, possibly indicative of an underlying systematic 
  uncertainty in the determination of the size evolution, of 
  differences in the selection function.  In order to take this 
  uncertainty into account, we assume as the observed value of 
  $\partial \log{\reff}/\partial z$ the mean between the values measured 
  by \citet{New++12b}, \citet{Dam++11}, \citet{Cim++12}, 
  \cite{Hue++13}, and we take their standard deviation as the 
  uncertainty: 
\begin{equation} 
\frac{\partial\log{\reff}}{\partial z} = -0.37 \pm 0.08. 
\end{equation} 
The mass dependence of $\reff$ is measured by \citet{New++12b} to be $\partial\log{\reff}/\partial \log{M_*} = 0.59 \pm 0.07$. 
The fit is done in a Bayesian framework. 
The posterior probability distribution for the gas fraction, as well as the redshift evolution of $\gamma'$ and $\reff$, is shown in \Fref{fig:post}. 
 
  The data prefer non-zero values of the gas fraction, with a
  median and $1-\sigma$ interval of $f_g = 0.08\pm0.01$. Purely dry
  merger models ($f_g = 0$) are disfavored at more than 99\% CL
  (formally at 8-$\sigma$).  The redshift dependence of $\gamma'$ is
  well matched by the model, and the $z$-dependence of $\reff$ is
  consistent with observations at the $2-\sigma$ level.  Although not
  plotted in \Fref{fig:post}, we verified that the dependences of
  $\gamma'$ on $\reff$ and $M_*$, as well as the dependence of $\reff$
  on $M_*$, are well consistent with observations.  This is expected,
  since the same observed scaling relations were used to initialize
  the mock sample at $z=0.3$. 
 
% {\bf As a test of the robustness of our results, we repeat the fit by
%  changing the parameters $a$ and $b$, which describe the variation in
%  $\gamma'$ due to dry mergers, to the values measured from
%  simulations with $\mhalo/M_* = 35$. The value of $f_g$ inferred in
%  this case is smaller than the value reported above, but pure dry
%  mergers are still excluded at more than $3-\sigma$.  Most galaxies
%  in our mock sample have $\mhalo/M_* > 35$. Since the decrease in
%  $\gamma'$ following dry mergers is stronger for larger values of
%  $\mhalo/M_*$, we conclude that for all reasonable models of dry
%  mergers involving massive ETGs, purely dissipationless megers fail
%  to reproduce observations.  }
 
%------------------------------------------------------------------------------ 
\section{Discussion}\label{sect:discuss} 
 
In \Sref{sect:comparison} we evaluated the mean evolution in the slope 
of the density profile $\gamma'$ of a mock sample of massive ETGs, 
under the assumption of growth by purely dry mergers.  We found that 
purely dry mergers produce a strong decrease in $\gamma'$, $d\gamma'/dz = 
0.33$ on average, inconsistent with observations. This result is 
robust against different choices for the SHMR and against variations 
in the values of the model parameters.  We then extended our model 
allowing for a modest amount of star formation in the mergers, and 
quantified the resulting effect on $\gamma'$. 
 
When considering simultaneously the evolution in density slope 
and size, we find that models with dissipation are strongly favored 
over purely dry merger models.  The most probable model has $f_g=0.08$, 
which according to our assumptions means that $8\%$ of the 
accreted baryonic mass consists of gas that falls to the center of the 
galaxy and forms stars.  The mock galaxies in our sample double 
their stellar mass between $z=1$ and $z=0$, on average. This would 
imply that $4\%$ of the final stellar mass of our galaxies being the 
result of in-situ star formation at $z<1$, or a specific star 
formation rate (sSFR) of $\sim0.01\mathrm{ Gyr}^{-1}$.  These numbers 
are consistent with the largest amount of recent star formation 
allowed by observations of ETGs, including spectral properties 
\citep{Tra++00a,Tre++02,Dame++09,Tho++10}, the evolution of the 
Fundamental Plane \citep{Tre++05}, UV \citep{Kav++11b} and mid-IR 
fluxes \citep{Fum++13} and spectral energy distribution fitting 
\citep{vDo++10,Ton++12}.  Thus we conclude that our model is as wet as it 
can be without violating known observational constraints.  The other 
extreme assumption of our model is that this cold gas falls all the 
way to the center before forming stars. Although this is clearly a 
toy-picture, it is at least qualitatively consistent with the ``blue 
cores'' seen in the center of massive ETGs at these redshifts 
\citep{MAE01,Tre++05,Pip++09a}. 
Observations of color gradients and their evolution \citep[e.g.][]{Szo++13} provide additional tests for the plausibility of the proposed scenario.
Preliminary calculations show the predicted change in colors due to wet mergers to be relatively small, and consistent with observations. 
However a more detailed comparison requires a careful assessment of the
observational selection function as well as additional assumptions on the 
star formation history and stellar populations.
This is beyond the scope of this paper and left for future work.

Purely dry mergers maximize the size growth of ETGs for a given 
increase in mass, and thus introducing some dissipation makes it 
harder to explain this observation.  Even though the tension with 
the size-growth data is much less than that between purely dry mergers 
and the evolution of the mass density profiles, it illustrates the 
challenges of achieving a fully self consistent and quantitative 
description of the evolution of massive early-type galaxies.  The 
discrepancy might be the result of progenitor bias 
\citep[e.g.][]{vdW++09,Cas++13}, which we have not accounted for. 
\citet{Lop++12} estimate that progenitor bias contributes $\sim20\%$ to the observed size evolution at $z<1$. 
This however raises the question of 
whether the observed evolution in the density slope might also be 
strongly influenced by progenitor bias effects.  Can the observation 
of $\left<\gamma'\right>\approx 2$ between $z=1$ and $z=0$ still be 
consistent with the dry merger model if we allow for the continuous 
emergence of new systems pushing the population average 
$\left<\gamma'\right>$ towards the measured value? This is very 
unlikely. Our model shows that dry mergers decrease the average 
density slope of a population of galaxies by $\sim0.3$ between $z=1$ 
and $z=0$.  On the other hand, the scatter in $\gamma'$ over the 
population of massive ETGs is as small as $0.12$ \citep{paperIV}.  In 
order to reproduce both the observed $\left<\gamma'\right>$ and 
scatter at $z=0$, the descendants of $z=1$ ETGs must be strongly 
outnumbered by newly born systems, at odds with observations 
\citep{Ilb++13,Cas++13}. 
 
The toy model developed in \Sref{sect:wet} is far from perfect, given 
the many simplifying assumptions it is based on. The effect of wet 
mergers on the density profile of ETGs is probably less pronounced in 
reality than in the idealized case considered here, and must be 
studied with dedicated numerical simulations in order to make 
quantitative statements. Nevertheless, our work shows with great 
clarity that i) dry mergers cannot be the {\it only} mechanism driving 
the evolution of massive ETGs; ii) a small amount of dissipation, 
consistent with observations, can bring the predicted evolution of 
$\gamma'$ in agreement with lensing measurements, as previously 
proposed by \citet{Ruf++11}. This finding is consistent with results 
from cosmological simulations that include dissipative effects, which 
predict trends in $\gamma'$ in qualitative agreement with the data 
\citep[][see 
\Fref{fig:gammap_evol_wwet}]{JNO++12,Rem++13,Dub++13}. 
 
\begin{figure} 
\includegraphics[width=\columnwidth]{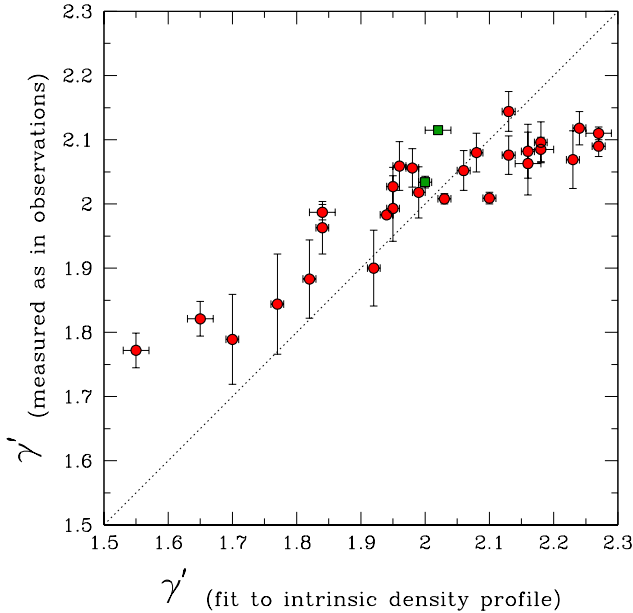} 
\caption{\label{fig:slope}  Total density slope $\gamma'$ as 
    defined in lensing and dynamics studies vs. $\gamma'$ obtained by 
    a direct fit to the angle-averaged density profile $\rho(r)$ for 
    the merger remnants (circles) and progenitors (squares) of the 
    $N$-body simulations of \citet{NTB09}. Vertical error bars account for 
    projection effects.} 
\end{figure}

%Alternative solutions have been proposed in the past. 
%\citet{Koo++06}, following theoretical findings \citep[e.g.][]{Gao++04,Deh05}, suggested major dry mergers between systems with equal $\gamma'$ as a way to preserve the density slope,  
As an alternative to the wet merger scenario, \citet{Bol++12}, based 
on the results of simulations by \citet{NTB09}, suggested off axis 
major dry mergers as a way to increase the density slope.   
  However, among the off-axis simulations of \citet{NTB09}, only those 
  with $M_*/\mhalo\sim 0.1$ produce remnants with $\gamma'$ higher 
  than in the progenitor, while $\gamma'$ decreases sharply in those 
  with the more cosmologically motivated stellar-to-dark mass ratio 
  $M_*/\mhalo\sim 0.02$.  Moreover, we stress that a crucial point in 
  connecting models with observations is defining $\gamma'$ in a 
  consistent way.  Theoretical works 
  \citep[e.g.][]{NTB09,JNO++12,Dub++13} often define $\gamma'$ by 
  fitting $\rho(r)$ with a power law $\rho(r) \propto r^{-\gamma'}$ 
  over a range of radii.  The two methods give different values of the 
  slope (see \Fref{fig:slope}) and we verified that, by measuring 
  $\gamma'$ as described in Section~\ref{sect:model}, the higher 
  values of $\gamma'$ predicted by the simulations of \citet{NTB09} 
  get revised downward by an appreciable amount.  The net effect, and 
one of the core result of our work, is that dry mergers decrease the density 
slope $\gamma'$, as defined in lensing and dynamics measurements.   
 
Finally, while our results suggest that dissipation plays a role in the evolution of ETGs, the origin of the cold gas involved in the process can still be subject of debate. 
Gas can for example be produced as a result of stellar mass loss. 
If internal sources of gas are the dominant ones, then wet mergers might not be necessary to keep the density profile from getting shallower, provided that the gas can effectively cool down and reach the central parts of the galaxy. 
Mergers might still play a role by inducing starbursts in the pre-existing gas. 
 
\section{Conclusions}\label{sect:concl}  
 
We studied the effect of dry mergers on the slope of the density 
profile of massive ETGs.  Both minor and major mergers produce a 
decrease in the density slope $\gamma'$, the effect being stronger for 
minor mergers, at fixed accreted mass. However, purely dry mergers 
produce a strong decrease in $\gamma'$ with time, inconsistent with 
lensing observations at more than $99\%$ significance. We thus 
developed a toy model to account for the infall of cold gas and star 
formation following wet mergers.  We found that it is sufficient to 
accrete 4\% of total mass in the form of cold gas to match the 
observed evolution in $\gamma'$ since $z\sim1$, while still be 
consistent with the observed size evolution.  We suggest a scenario 
where the outer regions of massive ETGs grow by accretion of stars and 
dark matter, while small amounts of dissipation and nuclear star 
formation conspire to keep the mass density profile constant and 
approximately isothermal. 
 
\acknowledgments  
We thank Philip Hopkins for useful comments and 
suggestions, Rhea-Silvia Remus for kindly providing predictions from cosmological simulations and the referee for constructive criticism that helped 
improve this paper.  AS acknowledges support by a UCSB Dean Graduate 
Fellowship.  TT acknowledges support by the Packard Foundation in the 
form of a Packard Research Fellowship. The SL2S project was supported 
by NASA through HST grants GO-10876, GO-11289, GO-11588. CN 
acknowledges financial support from PRIN MIUR 2010-2011, project ``The 
Chemical and Dynamical Evolution of the Milky Way and Local Group 
Galaxies'', prot. 2010LY5N2T. 
%------------------------------------------------------------------------------- 
% 
%\acknowledgments 
% 
% Boilerplate: 
%\input{acknowledgments2.tex} 
% 
%------------------------------------------------------------------------------- 
\bibliographystyle{apj} 
\bibliography{references} 
%------------------------------------------------------------------------------- 

\end{document}